\documentclass[
 prl,
 reprint,
 floatfix,
 amsmath,amssymb,
 showpacs,
 superscriptaddress,
 citeautoscript,
 nonatbib,
 frontmatterverbose
]{revtex4-1}

\usepackage{xr}
\makeatletter
\newcommand*{\addFileDependency}[1]{
  \typeout{(#1)}
  \@addtofilelist{#1}
  \IfFileExists{#1}{}{\typeout{No file #1.}}
}
\makeatother

\usepackage{amssymb}

\usepackage[breaklinks]{hyperref}
\usepackage[all]{hypcap}

\makeatletter
\renewcommand{\fnum@figure}[1]{\textbf{Fig.~\thefigure~}}
\makeatother

\hypersetup{
    plainpages=false,
    unicode=false,                          
    pdfborder={0 0 0},
    pdftoolbar=true,                        
    pdfmenubar=true,                        
    pdffitwindow=false,                     
    pdfstartview={FitH},                    
    pdftitle={},                            
    pdfauthor={},                           
    pdfproducer={LNCMI-CNRS-UGA-UPS-INSA},  
    pdfkeywords={High} {Magnetic} {Fields}, 
    pdfnewwindow=true,                      
    linktoc=section,
    colorlinks=true,                        
    linkcolor=blue,                         
    citecolor=red,                          
    filecolor=magenta,                      
    urlcolor=blue                           
}

\usepackage{graphicx}
\usepackage{float}
\usepackage{amsmath}
\usepackage{bm}
\usepackage{color, soul}
\usepackage[normalem]{ulem}
\usepackage{siunitx}
\usepackage{upgreek}
\usepackage[usenames,dvipsnames]{xcolor}
\usepackage{textcomp}
\usepackage{chemformula}
\usepackage[normalem]{ulem}
\usepackage{rotating}
\begin{document}

\title{Ultrafast magnetization induced by linearly polarized pulses is widespread in nonmagnetic semiconductors}

\author{Xiangzhou Zhu}
\email{xiangzhou.zhu@unitn.it}
\affiliation{Department of Physics, University of Trento, Trento, Italy}

\author{Junfeng Qiao}
\email{junfeng.qiao@epfl.ch}
\affiliation{Theory and Simulation of Materials (THEOS), and National Centre for
Computational Design and Discovery of Novel Materials (MARVEL), École Polytechnique Fédérale de Lausanne, Lausanne, Switzerland}

\author{Nicola Marzari}
\email{nicola.marzari@epfl.ch}
\affiliation{Theory and Simulation of Materials (THEOS), and National Centre for
Computational Design and Discovery of Novel Materials (MARVEL), École Polytechnique Fédérale de Lausanne, Lausanne, Switzerland}

\author{Matteo Calandra}
\email{m.calandrabuonaura@unitn.it}
\affiliation{Department of Physics, University of Trento, Trento, Italy}

\date{\today}

\begin{abstract}
Ultrafast optical on-off switching of magnetic order promises near-petahertz information processing. Recently, it has been proposed that non-magnetic semiconductors
with narrow band edges or strong exchange interactions could display ultrafast magnetization when photoexcited with linearly polarized femtoseconds pulses, but the experimental detection of this effect remains a challenge, mostly for the lack of suitable candidate compounds.
Here, we present a high-throughput first-principles screening of the MC3D database of experimentally known inorganic crystals, identifying nearly 440 non-magnetic semiconductors that develop spin polarization under photoexcitation with linearly polarized pulses via a light-induced exchange-driven instability.
We determine how the crystal field environment and band-edge orbital character govern
the magnitude and the type of magnetic order of the photoinduced
state and we unveil systematic chemical and periodic trends that provide intuitive guidance for materials selection. 
Our results argue that on-off switching of magnetization with linearly polarized femtosecond pulses is a widespread occurrence in non-magnetic semiconductors, opening novel avenues for experimental verification and application. 
\end{abstract}

 \maketitle


\section{Introduction}\label{sec1}
Ultrafast control of the spin degrees of freedom and of the magnetization in materials is a key requirement for quantum information, ultrafast memories and high-speed spintronics
\cite{lambert2014,walowski2016,kimel2019,fangPerspectivesLightControl2024}.
Femtosecond laser pulses have emerged as powerful tools to manipulate spin order at near-petahertz rates, as demonstrated experimentally by the  ultrafast demagnetization in Ni \cite{beaurepaire1996}, where the system is driven into 
a $\approx 30\%$ lower magnetic state in few ps.  
Even more challenging for information processing technology is light-induced ultrafast magnetization of a non-magnetic semiconductor; namely, the on and off switching of the magnetization of a nanodomain, going from a bit-zero configuration (no magnetization) to a bit-one state (finite magnetization) in a few hundred of femtoseconds (fs). 


A widely explored route toward light-induced magnetization in non-magnetic materials relies on circularly polarized light \cite{kimelUltrafastNonthermalControl2005,stanciuAllOpticalMagneticRecording2007}. 
The inverse Faraday effect \cite{vanderziel1965,pershan1966} provides a prototypical mechanism: circularly polarized light generates an effective magnetic field that couples to electronic spins through spin–orbit coupling (SOC), enabling helicity-dependent control of magnetic order \cite{stanciuAllOpticalMagneticRecording2007,radu2011,lambert2014,zhangAllopticalSwitchingMagnetization2022}. 
In contrast, linearly polarized light carries no net angular momentum, and therefore cannot induce magnetization through conventional angular-momentum transfer mechanisms. 
Nevertheless, theoretical and experimental works have demonstrated that linearly polarized light can also influence magnetic order through alternative pathways \cite{matsubaraUltrafastPhotoinducedInsulatorFerromagnet2007,siegristLightwaveDynamicControl2019,neufeldAttosecondMagnetizationDynamics2023}. However, the ultrafast magnetization of a non-magnetic insulator  with linearly polarized ultrafast pulses is considered to be quite rare; it has only recently been proposed in theoretical calculations of very few systems \cite{neufeldAttosecondMagnetizationDynamics2023,mariniTheoryUltrafastMagnetization2022}, and it has never been detected in experiments. In addition, clear guidelines are missing to identify promising materials.

For instance, calculations by Neufeld \emph{et al.} \cite{neufeldAttosecondMagnetizationDynamics2023} showed that  strong-field–driven electronic motion can generate transient spin and orbital magnetization through ultrafast charge redistribution and spin–orbit–coupled dynamics on sub-cycle timescales. Recent theoretical work by Marini {\it et al.} \cite{mariniTheoryUltrafastMagnetization2022} has proposed a Stoner-like \cite{stoner1938} route toward light-induced magnetization in non-magnetic semiconductors, where photoexcited carriers in a quasi-equilibrium electron-hole plasma can stabilize a transient magnetic state.
In this picture, photoexcitation redistributes carriers near the band edges, rather than supplying spin angular momentum directly, so that exchange splitting becomes energetically favorable, allowing a transient spin-polarized state to emerge in systems that are already close to an exchange instability in equilibrium.
Systems with large densities of states near the band edges, weakly dispersive bands, or intrinsically strong exchange interactions are therefore promising candidates.  
This identification suggests a systematic design strategy: rather than targeting strong SOC or helicity-dependent effects, one can instead look for semiconductors whose band-edge electronic structure favors exchange enhancement under photoexcitation, potentially expanding the range of non-magnetic semiconductors in which ultrafast magnetization by linearly polarized pulses can be detected.


In this work, we combine high-throughput first-principles screening with microscopic analysis to identify  non-magnetic semiconductors that can develop photoinduced magnetization through an exchange-driven, Stoner-like mechanism. 
We identify a broad class of compounds ($\approx 440$ ) having bandgap in the visible region that lie close to an exchange instability and become spin-polarized under photoexcitation. Surprisingly, we find that the ultrafast magnetization of a non-magnetic semiconductor with linearly polarized pulses is an extremely widespread phenomenon occurring in a large number of materials. 
We then rationalize these findings through structure–property motifs, showing how the crystal-field environment and band-edge orbital character govern the magnitude and magnetic order of the photoinduced state. 
Next, we uncover systematic chemical and periodic trends that provide intuitive guidance for materials selection. 
Finally, through representative case studies, we further illustrate the distinct microscopic mechanisms underlying the photoinduced magnetic response.
Together, these results move beyond isolated examples and establish a materials-level framework for discovering and designing semiconductors with exchange-driven photoinduced magnetization.

\section{Methods}

\subsection{High-throughput screening process}

To systematically identify candidate materials for light-induced magnetism, we constructed a high-throughput screening workflow starting from the Materials Cloud three-dimensional structure database (MC3D) \cite{huber2026} of experimentally known inorganic compounds, structures have been recalculated/relaxed using density-functional theory (DFT) at the PBEsol level, and carefully curated automated workflows.
Our objective is to focus on non-magnetic semiconductors that are electronically close to an exchange instability and may therefore develop spontaneous spin polarization under photoexcited carrier populations.

We first restrict the search to transition-metal containing compounds with at most 14 atoms in the primitive unit cell, enabling consistent first-principles treatment across a large dataset.
From this subset, we retained materials that are non-magnetic in their ground state and exhibit semiconducting character with an estimated PBEsol  \cite{perdew2008} Kohn-Sham band gap between 0.1 and 3 eV. 
These criteria target systems that are neither metallic nor have a too wide band gap, thereby ensuring a moderate optical excitation densities can realistically modify the electronic structure while preserving semiconducting behavior.
As generalized gradient approximations tend to underestimate the gap, this choice is a conservative estimate as the gap is expected to be larger.
This prescreening yields approximately 1700 materials.

To mimic ultrafast photoexcitation, we employ constrained density-functional theory (cDFT) \cite{marini2021} to impose quasi-equilibrium electronic occupations corresponding to a thermalized distribution of electrons and holes.
Specifically, electrons are removed from the valence band states and injected into the conduction band states, generating a fixed photoexcited carrier density of 0.2 electrons per unit cell (See Fig. S1  in Supplementary Material for discussion on robustness of this choice with respect to the other possible choice of fixing the number of electrons per unit volume).
The electronic structure is then recalculated self-consistently under these constrained occupations, allowing spontaneous spin polarization to emerge, if energetically favorable.
Spin–orbit coupling is not included at this stage, as the focus is on exchange-driven electronic instabilities rather than mechanisms relying on angular-momentum transfer from circularly polarized light.

Out of the 1700 compounds screened, approximately 440 develop finite magnetization under carrier excitation, indicating proximity to an exchange instability.
As mentioned, Kohn-Sham band gaps in generalized gradient approximation are known \cite{perdew1996} to systematically underestimate semiconductor gaps, especially in transition-metal compounds where correlation effects are significant. 
We therefore refine the band gaps of this subset of promising candidates beyond the initial estimates, which also underscores that the value of exact band gap is only relevant to the magnetization process in setting the actual energy of the photoexcitation.
Since there is no single database providing gap values calculated with more accurate functionals or the much more computationally expensive many-body perturbation theory, e.g. $G_0W_0$ approximation, for all compounds considered here, we adopt a hierarchical correction strategy.
Modified Becke–Johnson (MBJ) gaps are used when available \cite{choudharyComputationalScreeningHighperformance2018a};
otherwise, all-electron HSE06 values are employed \cite{nair2025}.
For the remaining materials, a machine-learning model trained on MBJ data is used to predict corrected gaps \cite{choudhary2021}.
This procedure provides consistent gap estimates while maintaining broad coverage across the dataset. From now on, we label as {\it corrected gaps} the gap values obtained with this more refined procedure.

\section{Results}

\subsection{Distribution of light-induced magnetization}
The high-throughput screening reveals that carrier-induced magnetization is not limited to a small number of special compounds, but instead emerges across a broad class of non-magnetic semiconductors.

\begin{figure*}[t]
    \centering
    \includegraphics[width=\linewidth]{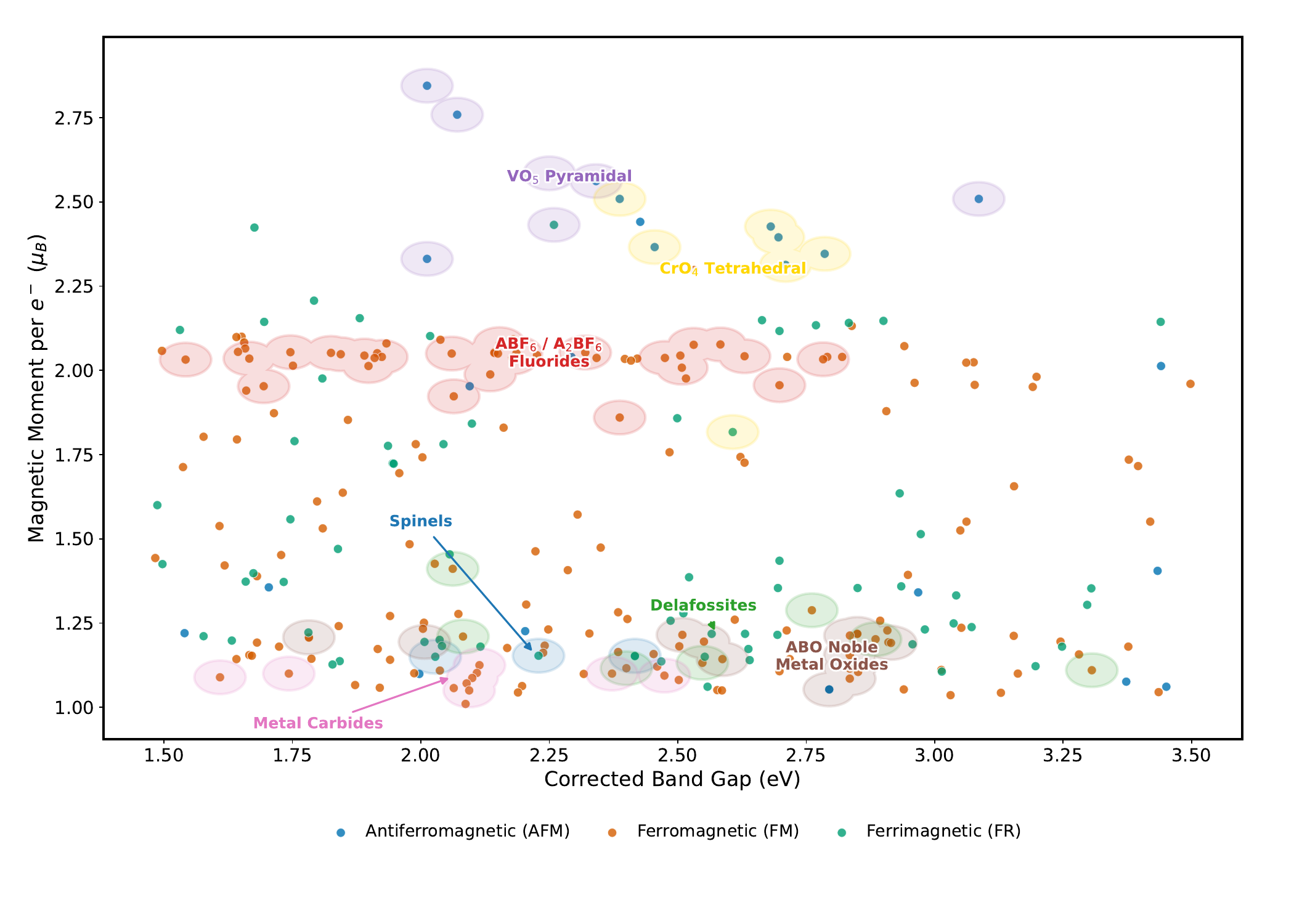}
    \caption{\textbf{Distribution of light-induced magnetization in non-magnetic semiconductors.} Magnetic moment per photoexcited electron as a function of corrected band gap for candidate materials identified in the high-throughput screening. Only compounds with corrected band gaps in the optically relevant range of 1.5–3.5 eV and absolute magnetic moments larger than 1 $\mu_B$ per photoexcited electron are shown, yielding around 280 materials. Blue, orange, and green symbols denote antiferromagnetic (AFM), ferromagnetic (FM), and ferrimagnetic (FR) photoexcited states, respectively. Characteristic material families are highlighed with different colors.}
    \label{fig:fig1_mag_gap}
\end{figure*}

Figure~\ref{fig:fig1_mag_gap} displays the absolute magnetic moment per photoexcited electron as a function of corrected band gap for materials within the experimentally relevant optical window of 1.5–3.5 eV and exhibiting an absolute magnetization larger than 1~$\mu_B$ per photoexcited carrier.
These thresholds identify about 280 materials that combine accessible optical gaps with a substantial exchange response under excitation.
The induced magnetic order is not restricted to a single spin configuration.
Antiferromagnetic (AFM), ferromagnetic (FM), and ferrimagnetic (FR) materials are represented by blue, orange, and green dots, respectively.
This diversity reflects that the resulting magnetic order depends on how exchange interactions reshape the spin splitting of the conduction and valence band edges in each compound, and will be discussed further in next Section.

The distribution of materials in the magnetization–gap map is clearly not random.
Two prominent dominant groups emerge in Fig. \ref{fig:fig1_mag_gap}: a group centered around approximately $2~\mu_B$ per photoexcited electron and a second group clustered near $1.2~\mu_B$ per electron, both spanning a broad range of band gaps.
The higher magnetic moment group is dominated by ABF$_6$, ABF$_4$ and A$_2$BF$_6$ fluorides, while spinels, delafossites, noble-metal oxides, and metal carbides cluster in the lower-moment regime.
This clustering of structural families in distinct regions suggests that the local coordination environment and the band-edge orbital character strongly influence the magnitude of the induced magnetic moment.
In addition, several AFM compounds exhibit magnetic moments exceeding $2~\mu_B$ per electron.
In these systems, the transition metal and anion carry opposite spin polarization, leading to large absolute moments despite compensated total magnetization.
Many of these materials belong to VO$_5$ pyramidal and CrO$_4$ tetrahedral coordination families, indicating that specific crystal-field geometries can strongly influence carrier-induced spin polarization.

Overall, these results show that exchange-driven photoinduced magnetization is not limited to a few isolated compounds, but emerges systematically across distinct families of transition-metal semiconductors.
The clear grouping of these families implies an underlying microscopic origin rooted in local coordination geometry and band-edge electronic structure. 
To uncover this connection, we next analyze the coordination motifs that govern the induced magnetic response.

\subsection{Coordination motifs}

The distribution of candidate materials in Fig.~\ref{fig:fig1_mag_gap} suggests that both the magnitude and the magnetic order of the photoinduced response are closely linked to the local coordination geometry of the transition-metal site. 
The coordination geometry fixes the crystal-field splitting \cite{orgel1960} of the metal $d$ orbitals (Fig~\ref{fig:fig2_crystal_fields}f), thereby selecting which orbital manifolds appear near the band edges. 
On the other hand, the chemical identities of the metal and ligand then control the degree of $p$--$d$ hybridization and band dispersion. 
Together, these factors determine the localization and interaction of the photoexcited electrons and holes, and therefore the exchange splitting and magnetic order that develop under photoexcitation. 
We therefore classify the candidate materials into five coordination motifs (Fig.~\ref{fig:fig2_crystal_fields}) and examine their electronic mechanisms in turn.

Before discussing the individual motifs, we first summarize the general principles that govern the magnetic order of the photoexcited state. 
The magnetic order is governed not only by the dominant atomic projection of the band edges, but also by whether the electron and hole spin splittings are coupled through the same localized structural unit or through different sublattices.
When a photoexcitation transfers carriers between hybridized anion $p$ valence states and metal $d$ conduction states on different sublattices, the induced spin polarizations at the two band edges often develop with opposite signs, favoring AFM or FR responses. 
By contrast, when both band edges arise predominantly from the same metal-centered crystal-field manifold, or when one band edge is much more exchange-active and the other follows its exchange field, the spin splitting tends to develop with the same sign, favoring FM order.
With this distinction in mind, we now discuss the five coordination motifs indentified.

\subsubsection*{Octahedral coordination motifs}

The first motif corresponds to transition-metal ions in an octahedral coordination environment, where the crystal field splits the $d$ orbitals into lower-energy $t_{2g}$ and higher-energy $e_g$ states (see Fig. ~\ref{fig:fig2_crystal_fields} (f)). 
The occupation of these orbitals determines the character of the band edges and, consequently, the magnitude and nature of the carrier-induced magnetic response.

A first class is that of octahedral $d^0$ compounds (in purple in Fig.~\ref{fig:fig2_crystal_fields} (a)), where the valence band maximum (VBM) is typically dominated by anion $p$ states and the CBM by metal $t_{2g}$ states. 
Photoexcitations therefore transfer carriers between different sublattices. 
Because the conduction $t_{2g}$ states are often relatively dispersive, the exchange instability is usually stronger on the valence-band side, giving induced moments of $\sim 1.2~\mu_B$ per photoexcited electron. 
The resulting spin polarizations at the VBM and CBM tend to develop opposite signs and stabilize ferrimagnetic order. 

The second class corresponds to the fully occupied $d^{10}$ octahedral compounds (in {blue in Fig.~\ref{fig:fig2_crystal_fields}  (a)). 
In this case, the VBM typically consists of hybridized metal $d$ and anion $p$ states, while the CBM is more dispersive and often has metal $s$ character. 
As a result, the induced magnetization is mainly driven by the localized holes at the valence band edge, while the more delocalized conduction electrons follow the exchange field generated by these polarized holes.
These systems therefore typically show ferromagnetic photoexcited states with more moderate induced moments of the order of $1.2~\mu_B/e^-$. 

A stronger magnetic response appears in low-spin $d^6$ compounds
(in {yellow in Fig.~\ref{fig:fig2_crystal_fields} (a)), where the occupied $t_{2g}$-like states form VBM and the empty $e_g$-like states form the CBM. 
In this case, photoexcitations transfer carriers between crystal-field-split states localized on the same metal-centered unit, so that the photoexcited electron and hole align parallel and favor the ferromagnetic response.
However, the absolute magnitude of this induced moment usually depends on the flatness of the $e_g$ conduction band, which is dictated by the degree of metal $d$ and ligand $p$ hybridization.
Especially in fluorides the deep $p$ states reduce these $p$-$d$ hybridization, preserving the narrow $e_g$ band. 
Therefore, the exchange splitting dominates under photoexcitation, allowing the spin polarization to approach the theoretical maximum of $2~\mu_B$ per electron. 
Conversely, in coordination environments with stronger $p$-$d$ mixing, the $e_g$ states become more dispersive, leading to a weaker induced magnetic moment.

Overall, the octahedral motif highlights three distinct electronic scenarios for light-induced magnetization: 
via $p\rightarrow d$ charge transfer in $d^0$ compounds; with hole-dominated magnetic response in $d^{10}$ materials; and with a stronger localized $d\rightarrow d$ excitations in $d^6$ systems. 
Together, these cases demonstrate that while the underlying physics remains constant, the magnitude and character of the magnetic response are ultimately controlled by the specific crystal-field-derived band-edge states and their spatial localization.

\subsubsection*{Tetrahedral coordination motifs}

The second motif corresponds to transition metals in a tetrahedral coordination environment. 
In this geometry, the crystal-field splitting of the metal $d$ orbitals is reversed relative to the octahedral case, with the $e$ manifold lying below the $t_2$  (see Fig.~\ref{fig:fig2_crystal_fields} (f)).

In tetrahedral $d^0$ compounds, the VBM is dominated by anion $p$ states, while the CBM is derived from the empty metal $e$ states.
Because photoexcitations transfer carriers between these states on distinct sublattices, the induced spin polarizations at the two band edges develop with opposite signs, stabilizing a ferrimagnetic (FR) order with induced moments of the order of $1.2~\mu_B$ per photoexcited electron.
Moreover, the microscopic driver of this FR state differs from its octahedral counterpart: 
while octahedral $d^0$ systems often exhibit a valence-band-driven exchange instability, the tetrahedral $d^0$ magnetic response is predominantly conduction-band-driven with relatively narrow  metal $e$ states.

A notable exception occurs in certain low-dimensional compounds containing CrO$_4$ tetrahedra (in {purple in Fig.~\ref{fig:fig2_crystal_fields}  (b)). 
In these materials, the states at both band edges are sufficiently localized to sustain strong exchange splitting after photoexcitation. 
The resulting spin polarization develops with opposite signs on the oxygen- and chromium-centered components, leading to AFM configurations with large absolute magnetization. 

Another class is given by $d^{10}$ halides, including GaCuCl$_4$, GaCuBr$_4$, and AlCuCl$_4$ ({yellow in Fig.~\ref{fig:fig2_crystal_fields}  (b)). 
Here, the transition metal $d$ shell is fully occupied, and the valence band typically consists of hybridized metal $d$ and anion $p$ states, while the conduction band may involve more dispersive $s$-derived states. 
These materials generally exhibit ferromagnetic ordering with comparatively small induced magnetic moments, driven by laser-induced valence-band electron depletion.

\begin{figure*}
    \centering
    \includegraphics[width=\linewidth]{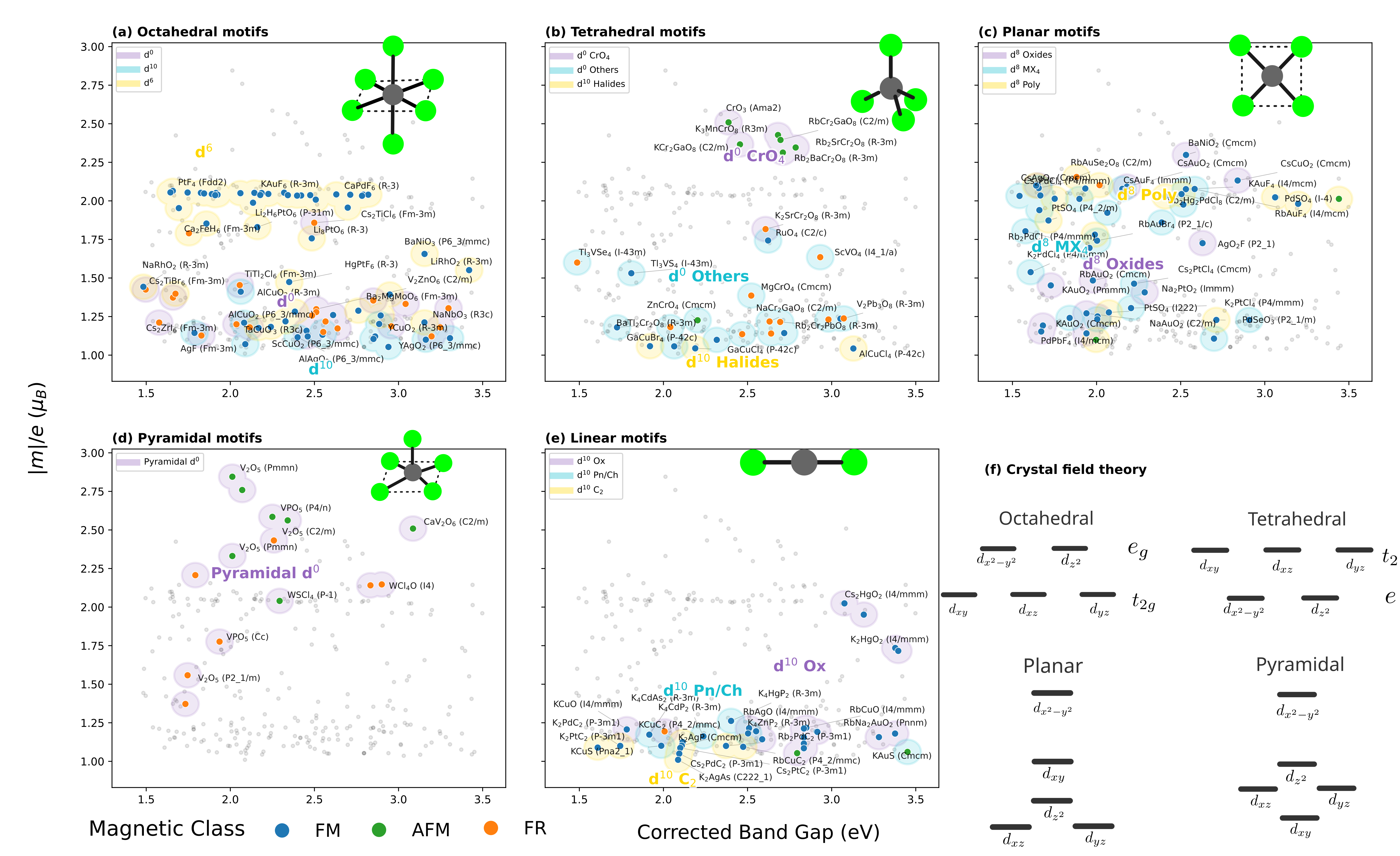}
    \caption{\textbf{Structure–property motifs for light-induced magnetization.} Magnetic moment per photoexcited electron versus corrected band gap, classified by local coordination motif: (a) octahedral, (b) tetrahedral, (c) planar, (d) pyramidal, and (e) linear. Insets show schematic crystal-field environments for each motif. Colored symbols indicate the magnetic class of the photoexcited state (FM, AFM, and FR), while gray points represent the full candidate distribution shown for reference. They reproduce the same set of nearly 280 materials shown in Fig. \ref{fig:fig1_mag_gap}. The colored blobs highlights the chemically and electronically related subclasses within each coordination motif, as shaded areas indicated by the panel legends.}
    \label{fig:fig2_crystal_fields}
\end{figure*}

\subsubsection{Planar coordination motifs}
All planar compounds identified in our screening share a common $d^8$ electronic configuration under a square-planar crystal field. 
In this geometry the $d_{x^2-y^2}$ orbital is pushed to higher energies (see Fig.~\ref{fig:fig2_crystal_fields}(f)) forming a strongly antibonding state with ligand $p_\sigma$ orbitals. 
Consequently, the CBM is mainly derived from this hybridized $d_{x^2-y^2}$ state, while the VBM typically consists of weakly antibonding ligand $\pi$ states mixed with other occupied metal-$d$ states.

The planar materials in our dataset include extended oxide networks (such as edge-sharing ABO$_2$ compounds), isolated square-planar halides (ABX$_4$), and polyatomic-ligand environments (such as PdSO$_4$), shaded in purple, blue and yellow in Fig.~\ref{fig:fig2_crystal_fields} (c) respectively. 
Across these families, the magnitude of the light-induced magnetization is governed by the dispersion of the $d_{x^2-y^2}$ conduction band.
Weaker metal--anion overlaps and longer bond lengths narrow this band, enhance the density of states at the CBM, and promote stronger exchange splitting. 
Even subtle structural variations can tune the competition between kinetic energy and exchange energy.
For example, the larger lattice constant in CsAuO$_2$ relative to KAuO$_2$ reduces inter-site hopping of the $d_{x^2-y^2}$ states, resulting in a flatter conduction band and a larger induced moment ($\approx 2~\mu_B$ in CsAuO$_2$ and $\approx 1.2~\mu_B$ in KAuO$_2$).

Interestingly, most planar compounds stabilize in a ferromagnetic configuration, even though the VBM and CBM often have different orbital character. 
In the square-planar $d^8$ motif, the ligand $p_\pi$ valence states are not the direct bonding counterpart of the conduction-band $d_{x^2-y^2}$--$p_\sigma$ manifold, and the photoexcited holes do not drive an opposite spin polarization of the conduction electrons as often happens in the $d^0$ case. 
Instead, the planar $d^8$ systems are closer to the low-spin $d^6$ octahedral case where the two band edges tend to acquire spin splitting with the same sign, leading to a ferromagnetic photoexcited state.

\subsubsection{Pyramidal coordination motifs}

The pyramidal compounds in our dataset are predominantly $d^0$ systems, typically based on the transition metals V$^{5+}$, Mo$^{6+}$, and W$^{6+}$. 
Unlike $d^0$ octahedral and tetrahedral systems, where the CBM is derived from the $t_{2g}$ and $e$ manifolds respectively, pyramidal compounds exhibit a different orbital hierarchy. 
In these pyramidal geometries, the strong axial metal–oxygen bond introduces a pronounced anisotropy.
Consequently, the largely non-bonding $d_{xy}$ state determines the CBM, while the VBM remains dominated by oxygen p-states.

In particular, VO$_5$ based systems (such as V$_2$O$_5$, CaV$_2$O$_6$, and VOPO$_4$) favor antiferromagnetic ordering and lead to opposite spin polarization on V and O sites, similar to the CrO$_4$ tetrahedral systems (see Fig. Fig.~\ref{fig:fig2_crystal_fields} (d)). 
Consequently, while the global magnetic order is AFM, the absolute local moments on the respective atomic sublattices become remarkably large, exceeding 2 $\mu_B$ per photoexcited electron.
MOX$_4$ systems (M = Mo, W) follow a similar structural motif, but the more extended $4d$ and $5d$ orbitals broaden the conduction band and reduce the exchange splitting, yielding weaker induced local moments and typically FR order.

\subsubsection{Linear coordination motifs}

The last structural motif that emerges is that of closed-shell $d^{10}$ compounds with linear coordination geometries. 
This motif includes three chemical subclasses, namely oxides, pnictides/ chalcogenides, and compounds containing C$_2$ ligands, which are highlighted in purple, blue, and yellow in Fig.~\ref{fig:fig2_crystal_fields} (e), respectively.
In these materials, the CBM is highly dispersive, while the VBM is typically formed by hybridized metal $d$-anion $p$ states and leads to exchange instability after photoexcited. 

Within this motif, exchange splitting is primarily driven by the holes in the narrow valence band edge. 
Unlike the $d^0$ systems, where electrons and holes occupy different sublattices and tend to develop opposite spin polarization, the more dispersive conduction states here resist forming an independent spin splitting.
Instead, they acquire a spin splitting following the exchange field generated by the polarized holes, thereby stabilizing mainly ferromagnetic order.
Therefore, most compounds exhibit induced magnetic moments close to $1.0$–$1.2~\mu_B$ per photoexcited electron, except for certain A$_2$HgO$_2$ compounds.


\subsection{Chemical and periodic trends}
While the coordination motif primarily determines the orbital character at the band edges, the chemical composition provides an additional perspective on the distribution and magnitude of the induced magnetic response. 
To place the motif-based analysis in a broader chemical context, Fig.~\ref{fig:stats_trend} summarizes elemental statistics for the full set of 440 candidate materials exhibiting light-induced magnetization.

Figure~\ref{fig:stats_trend}a presents the frequency distribution of anion species. 
The bars denote the number of materials containing a given anion, while the black curve indicates the mean induced magnetic moment per photoexcited electron, with error bars representing the standard deviation. 
Oxides appear most frequently among the identified candidates, consistent with their prevalence in the database and their ability to support strong metal–oxygen hybridization. 
Fluorides has the second most counts and the highest average light induced magnetic moments, approaching $2~\mu_B$ per electron. 
This behavior is consistent with the strong exchange enhancement observed in ABF$_6$ and A$_2$BF$_6$ families, with a relatively localized spatial distribution of fluorine $p$ orbitals.

Figure~\ref{fig:stats_trend}c further illustrates periodic trends across rows of the periodic table. 
Within each anion family, elements in lower periods (N, O, F) display larger induced moments, consistent with increased orbital localization in the smaller $p$ orbitals.
This trend reflects the increased orbital localization and reduced band dispersion associated with lighter $p$-block elements, which enhances exchange splitting near the band edges. 
In contrast, heavier anions possess more spatially extended $p$ orbitals, leading to broader bands and a reduced tendency toward exchange instability.

Figure~\ref{fig:stats_trend}b shows the corresponding distribution for transition-metal species, following the same convention as in panel~\ref{fig:stats_trend}a. 
Late $4d$ and $5d$ elements, particularly Pd, Pt, and Au, appear most frequently among the identified candidates. 
These elements can not only commonly realize closed shell $d^{10}$ configurations, but can also stabilize $d^{6}$ or $d^{8}$ occupancies, which are favorable for light-induced spin polarization in octahedral and planar coordination motifs.

Figure~\ref{fig:stats_trend}d further groups the transition-metal statistics by period row ($3d$, $4d$, and $5d$ series), revealing a systematic decrease in the mean induced magnetic moment as we move down the groups. 
This trend can be attributed to the spatial localization of the orbitals. 
The strong localization of $3d$ orbitals enhances electron correlation and exchange splitting, thereby strengthening the photoinduced spin polarization. In contrast, the more spatially extended $4d$ and $5d$ orbitals lead to broader bandwidths and reduced exchange enhancement, resulting in a smaller average magnetic response. 
Within the $3d$ series, V, Co, and Ni exhibit stronger magnetization. 
As discussed previously, this behavior is related to  their specific structural preferences, namely the highly anisotropic vanadyl pyramidal motif for $V^{5+}$ and the $d^6$ octahedral motif for Co$^{3+}$ and Ni$^{4+}$.

\begin{figure*}
    \centering
    \includegraphics[width=\linewidth]{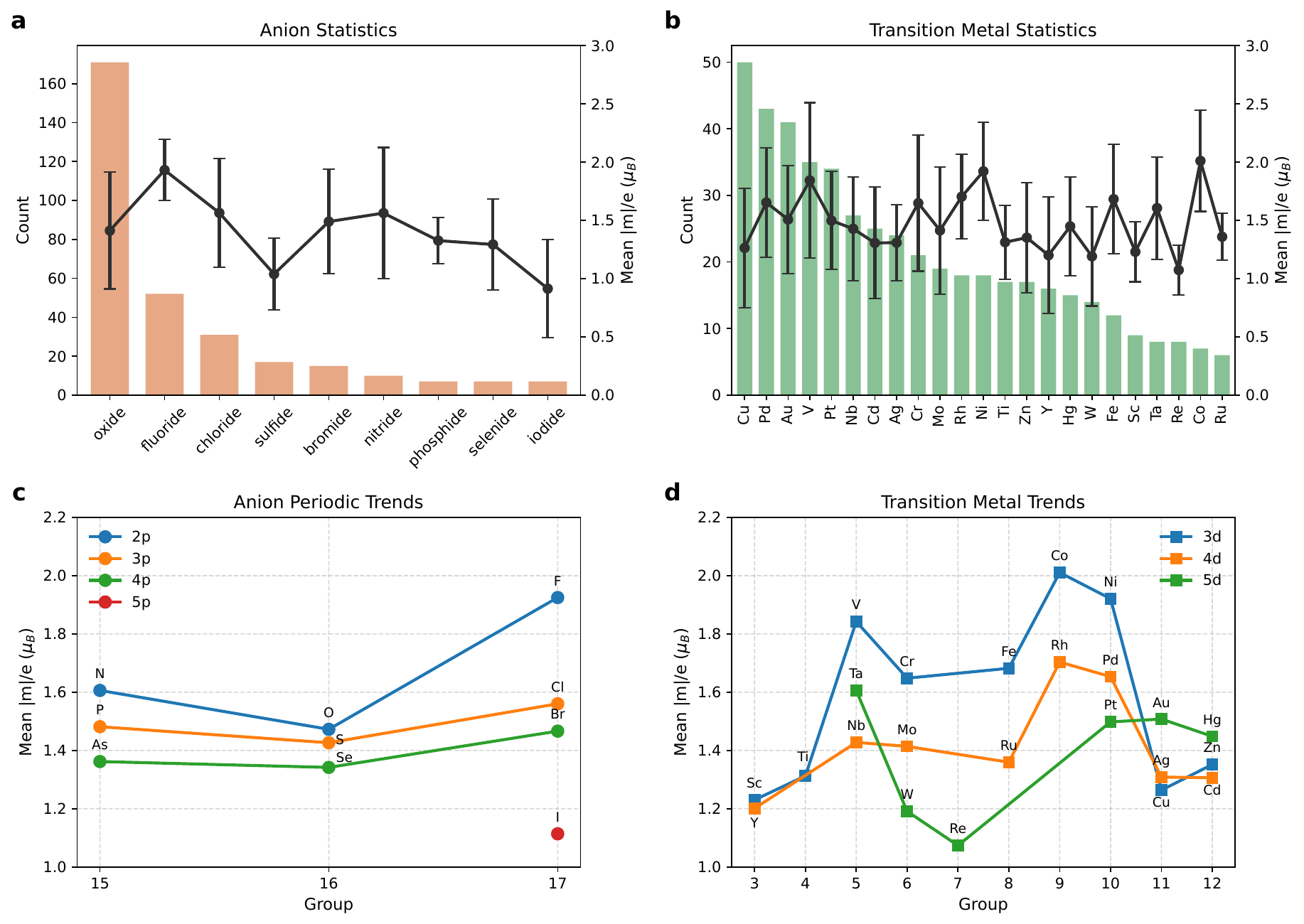}
    \caption{\textbf{Chemical and periodic trends of the candidates identified. } Statistics of anion species (a)  and transition-metal species (b) in the magnetic candidate set. Bars show the number of candidate materials containing each anion, while the black line gives the mean induced magnetic moment per photoexcited electron. Error bars indicate the standard deviation.  
    Mean induced magnetic moment grouped by anion (c) and transition-metal species (d) family and periodic row.}
    \label{fig:stats_trend}
\end{figure*}

\subsection{Representative materials}
To further illustrate the physical mechanisms of the light-induced magnetization, we discuss three representative compounds: a tetrahedral d$^0$ oxide CrO$_3$, the $d^{10}$ compound with linear coordinate motifs CuAlO$_2$, and the $d^6$ octahedral fluoride Cs$_2$PdF$_6$.
These materials span different crystal field environments and band edge orbitals so that we can understand the microscopic origin of the photoinduced magnetic instability. The complete set of compounds displaying light-induced ultrafast magnetization  is available in the Supplemental Materials.

\begin{figure*}
    \centering
    \includegraphics[width=\linewidth]{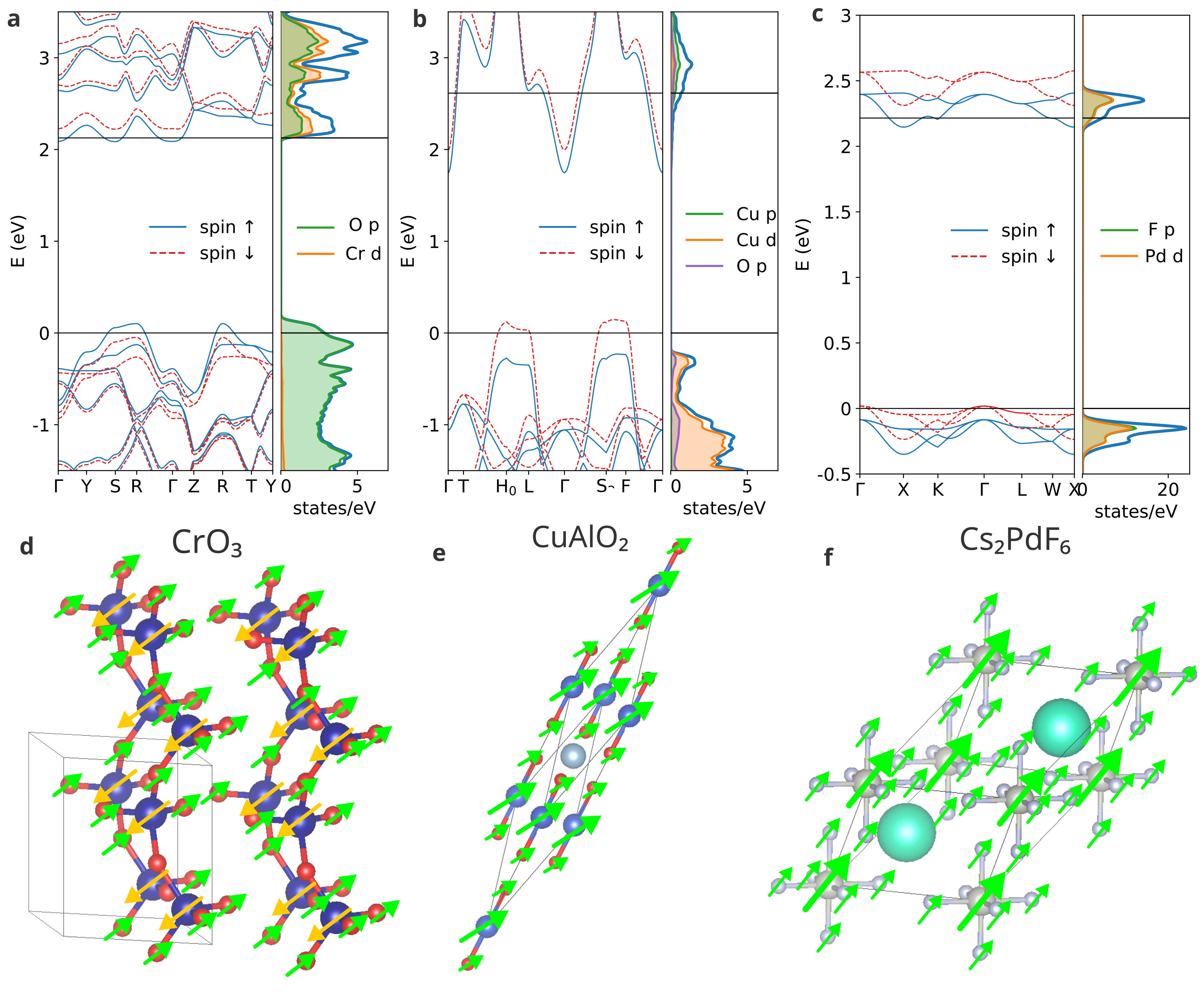}
    \caption{\textbf{Representative microscopic mechanisms of light-induced magnetization.} Spin-resolved band structures of the photoexcited state and projected density of states (PDOS) for spin up for (a) CrO$_3$, (b) CuAlO$_2$, and (c) Cs$_2$PdF$_6$, together with the corresponding structure and collinear spin vectors in (d), (e), and (f), respectively. The spin down DOS is similar to the spin up one, simply shifted in energy.}
    \label{fig:fig4_represent}
\end{figure*}

\subsubsection{CrO$_3$}
Chromium trioxide CrO$_3$ forms 1D chain-like phase with the $Ama2$ space group \cite{stephens1970}.
In this compound, Cr is in a $d^0$ configuration and the structure consists of corner-sharing CrO$_4$ tetrahedra units as shown in Fig.~\ref{fig:fig4_represent}d.

The band structure of structurally relaxed photoexcited structure is shown in Fig.~\ref{fig:fig4_represent}a. 
The influence of this structural optimization is analyzed in Supplementary Material Fig. S2.
As shown in the projected density of states (PDOS) on the right, the valence band maximum (VBM) is dominated by non-bonding O $2p$ states, while the conduction band minimum (CBM) is mainly formed by the Cr $3d$ $e$ states with O $2p$ admixture.
In the tetrahedral crystal field of the CrO$_4$ units, the $e$ states are pushed below the $t_2$ manifold and define the conduction edge (see Fig.~\ref{fig:fig2_crystal_fields}f).

Upon photoexcitation, electrons populate these relatively localized Cr $e$ states, leads to exchange splitting and stabilizes a spin-polarized state (spin up in the figure). 
Because the CBM is mainly Cr 
$3d$ whereas the VBM is dominated by O $2p$ states, photoexcitation creates electrons and holes on different atomic sublattices.
The O $p$ valence states and Cr $d$ conduction states are coupled through Cr–O covalency, which leads to opposite spin polarization on the O- and Cr-centered components, as seen from the opposite spin splitting of the VBM and CBM of Fig~\ref{fig:fig4_represent}a.
Therefore, the photoinduced state shows an antiferromagnetic real-space character.

This behavior is closely related to the mechanism proposed for V$_2$O$_5$ by Marini \textit{et al.} \cite{mariniTheoryUltrafastMagnetization2022}.
More generally, it can drive antiferromagnetic or ferrimagnetic photoinduced states in materials with occupied $p$ in VBM and unoccupied $d$ in CBM ($d^0$), specifically those containing octahedral, tetrahedral or pyramidal local coordinates.

\subsubsection{CuAlO$_2$}
Delafossite  CuAlO$_2$ is representative of the $d^{10}$-type oxides in our screening.
It crystallizes in the rhombohedral $R\bar{3}m$ space group, consisting of alternating stacks of linearly coordinated O--Cu--O units and two-dimensional layers of edge-sharing AlO$_6$ octahedra, as shown in Fig.~\ref{fig:fig4_represent}e.

The band structure of the relaxed photoexcited state is shown in Fig.~\ref{fig:fig4_represent}b. 
The VBM is dominated by Cu $3d$ states, with O $2p$ hybridization, that exhibits a characteristic Mexican-hat-like dispersion, whereas the CBM is more dispersive and  mainly of Cu $p$ and O $p$ character. 
Thus, the photoexcited electrons and holes are not strongly localized on two sublattices with opposite spin response, like in the $d^0$ case of CrO$_3$.
Instead, magnetization is primarily driven by holes in the narrow Cu $d$ valence manifold.
The more delocalized conduction electrons subsequently follow the exchange field generated by these spin-polarized holes.
As a result, the VBM and the CBM develop spin splitting of the same sign, as shown in Fig.~\ref{fig:fig4_represent}b, leading to a photoinduced ferromagnetic response.

Interestingly, CuAlO$_2$ has also been proposed to exhibit Stoner ferromagnetism under hole doping \cite{iordanidou2021}, which is consistent with the dominant role of the valence-band holes found here. 
This example illustrates that systems with fully occupied $d$ states at the valence-band edge, spanning linear, tetrahedral, and octahedral coordination environments, provide a distinct route to light-induced magnetization. 
The induced magnetization of these systems are ferromagnetic around 1.2 $\mu_B$ per photoexcited carrier, this value reflects that the magnetic instability is driven almost entirely by the localized holes in the valence band, rather than by strong hybridization of both band edges as in CrO$_3$.

\subsubsection{Cs$_2$PdF$_6$}
Cs$_2$PdF$_6$ is representative of the $d^6$ fluoride family, which forms one of the groups with the highest magnetic response in our screening. 
Crystallizing in the highly symmetric cubic $Fm\bar{3}m$  space group, it adopts a vacancy-ordered double perovskite structure.
This geometry consists of isolated octahedral PdF$_6$ units separated by ordered B-site vacancies and interstitial Cs cations, as shown in Fig.~\ref{fig:fig4_represent}f. 
In this strong octahedral crystal field, the Pd$^{4+}$ cation exhibits a low-spin d$^6$ configuration.
The lower-energy $t_{2g}$ orbitals are fully occupied while the higher-energy $e_g$ orbitals remain empty, leading to a non-spin-polarized ground state.

The relaxed band structure is shown in Fig.~\ref{fig:fig4_represent}c. 
Both the VBM and CBM are strongly hybridized Pd-$d$ and F-$p$ states of the isolated PdF$_6$ octahedron.
The VBM has mainly occupied t$_{2g}$-like Pd–F character, while the CBM has empty e$_g$-like antibonding Pd–F character.
Due to the compact nature of the F $2p$ orbitals, orbital overlap is limited, leaving the band edges weakly dispersive compared to Cl or Br.
This localization makes this family particularly favorable for light-induced magnetization.
Upon photoexcitation, electrons populate these weakly dispersive Pd $e_g$ states and develop exchange splitting. 
Unlike the charge transfer transition in CrO$_3$, the two band edges here arise from the same Pd $d$ orbitals.
Because both band edges are localized on the same PdF$_6$ octahedral unit and remain weakly dispersive, both photoexcited holes and electrons can participate in exchange splitting.
Their spin splittings develop with the same sign, leading to a large FM response close to 2 $\mu_B$ per photoexcited carrier.

A similar localized $d$--$d$ excitation process can occur in $d^8$ systems with square-planar coordination, where the VBM has mainly $t_2$ character and CBM $e$ character.
This mechanism spans in several material classes, including A$_2$BF$_6$ vacancy ordered double perovskites, ABF$_6$ isolate octahedral fluoride, and ABF$_4$ planar materials.
These compounds form a group that exhibits a light induced magnetic moment of nearly 2 $\mu_B$ per photoexcited carrier, as shown in Fig.~\ref{fig:fig1_mag_gap}. 

\subsection{Promising candidates for experimental verification}

Our study also identifies many promising candidate compounds for subsequent experimental verification (see Supplemental Materials for a complete list, which includes Refs. \citenum{vansetten2018,choudharyComputationalScreeningHighperformance2018a,pellicer-porres2006}). 
In selecting experimentally relevant materials, both the magnitude of the predicted photoinduced magnetic response and the feasibility of optical excitation should be considered. 
Among the highest-moment candidates, low-dimensional vanadium-based compounds are particularly attractive, with induced magnetic moments exceeding $2~\mu_B/e$. 
Representative examples include VOPO$_4$ (P4/n) \cite{jordan1973}, CaV$_2$O$_6$ (C2/m), and MgV$_2$O$_6$ (C2/m) \cite{jin2022}, all of which feature VO$_5$ pyramidal motifs that support large AFM or FR local spin polarization upon photoexcitation.

Beyond this vanadium family, light-induced magnetization is also predicted in several widely studied functional materials, albeit with more moderate responses of about $1.3$--$1.9~\mu_B/e$. 
These include perovskite-related compounds such as the ferroelectric oxide NaNbO$_3$ (R3c) \cite{mishra2007} and the vacancy-ordered halide perovskite Cs$_2$TiCl$_6$ (Fm$\bar{3}$m), which has been investigated as a lead-free optoelectronic semiconductor \cite{kong2020}. 
Other notable candidates include the classic delafossite transparent conducting oxide CuAlO$_2$ (R$\bar{3}$m) \cite{kawazoe1997} and Tl$_3$VSe$_4$ (I$\bar{4}$3m), a compound of interest for its ultralow thermal conductivity and thermoelectric-related properties \cite{mukhopadhyay2018}.

For these representative candidates, we provide in the Supplementary Materials the electronic band structures and density of states of the photoexcited magnetic states, together with the ground-state imaginary part of the dielectric function, $\mathrm{Im}\varepsilon(\omega)$. 
The former characterize the spin-polarized electronic structure upon photoexcitation, whereas the latter provides a crude estimate of the energy ranges of the dipole-allowed optical excitation along the three spatial directions within the independent-particle approximation.

\section{Discussion}
In this work, we identify a broad set of nearly 440 candidate non-magnetic semiconductors for ultrafast magnetization   driven by linearly polarized fs pulses.
By combining high-throughput screening of known inorganic compounds from the MC3D database with motif analysis, we show that this effect is not limited to a few exceptional compounds, but arises across multiple classes of materials with suitable band-edge electronic structures, and it is a very common occurrence. 
Moreover, by combining crystal-field-based motif classification and representative case studies, we develop a microscopic understanding of this behavior and identify the electronic factors that govern the magnitude and magnetic order of the induced magnetic moments.
Our results reveal two favorable situations for generating a strong light-induced magnetic response. 

The first occurs in compounds where both photoexcited electrons and holes occupy localized band-edge manifolds within the same metal--ligand coordination unit. 
This situation is realized in several $d^6$ and $d^8$ systems, especially in fluoride families such as ABF$_6$ and A$_2$BF$_6$. 
In these compounds, the relevant band edges arises due to  strong metal-Fluorine hybridization. 
The compact F $2p$ orbitals reduce inter-unit hopping and preserve narrow band dispersions near the gap. 
This enhances the density of states and allows both electron and hole to contribute to the exchange splitting, producing large ferromagnetic moments that approach $2\,\mu_B$ per photoexcited carrier.

The second favorable situation occurs in low-dimensional V- and Cr-oxides, where both the O-$p$-derived valence edge and the metal-$d$-derived conduction edge are sufficiently localized to support sizable exchange splitting upon photoexcitation.
The induced spin polarization therefore develops on different sublattices and can acquire opposite signs on the metal- and oxygen-centered states. 
This leads to antiferromagnetic photoinduced order and can yield large absolute moments exceeding $2\,\mu_B$ per photoexcited carrier.

Moreover, our results show that the magnetic order of the photoinduced state is determined primarily by the orbital character of the two band edges.
When photoexcitation transfers carriers between the anion $p$ valence states and the metal $d$ conduction states, the induced spin polarization on the two sublattices tends to align oppositely, favoring AFM or FR responses.
By contrast, FM order is favored when both band edges are the same metal-centered $d$ states, or when the valence edge and the conduction edge are not the bonding-antibonding counterpart, as in linear coordination motifs. 

Beyond the microscopic mechanism, our results also show that the most promising candidates occupy a favorable band-gap regime for optical control. 
The light-induced magnetic response remains robust across a broad range of corrected gaps, with around 200 compounds combining substantial induced moments above 1 $\mu_B$ per photoexcited electron and corrected band gaps in the experimentally relevant 1.5-3.5 eV window. 
This indicates that the carrier-driven Stoner instability identified here is compatible with realistic optical excitation energies rather than being restricted to a few rare semiconductors.
Furthermore, the physical picture is further supported by the earlier identification of doped-induced itinerant magnetism in nonmagnetic semiconductors, including CuAlO$_2$ \cite{iordanidou2021}, V$_2$O$_5$ \cite{xiao2009}, FeS$_2$ \cite{lei2021} and arsenopyrites such as FeAsS \cite{leiComputationalSearchItinerant2022}. 
Notably, these materials are found in our ultrafast magnetization database (see Supplementary Material).

Still, it is useful to recall some several limitations of the present approach. 
First, spin-orbit coupling is not included in the screening, and this may influence the quantitative results, especially for $5d$ compounds.
Second, calculated band gaps are not fully quantitatively accurate, so absolute gap values and corresponding optical thresholds should be treated with caution.
Last, our search is limited to bulk materials in the MC3D database and does not include two-dimensional materials \cite{mounet2018,campi2023}, which could host additional promising candidates.

In conclusion, we carried out a high-throughput search for light-induced magnetization in nonmagnetic semiconductors and uncovered a broad range of candidate materials with sizable photoinduced magnetic moments. 
Our analysis shows that this effect can emerge across diverse material classes through a carrier-driven exchange-splitting mechanism, rather than being restricted to a few isolated compounds. 

By combining motif classification with representative case studies, we further established a microscopic understanding of the phenomenon and showed that both the magnitude and magnetic order of the induced moment are controlled by the electronic structure at the band edges. 
Taken together, these results demonstrate the generality of light-induced magnetization in semiconductors and open a route toward the discovery and design of materials for optical control of magnetic properties.

\section{Computational Methods}
\subsection{Database and screening workflow}
We started from the PBEsol-v2 dataset in the MC3D database \cite{huber2026} of experimentally known bulk stoichiometric inorganic compounds. 
Only unique phases are present, and the DFT pre-relaxed MC3D structures were used directly. 
The screening was restricted to compounds containing $3d$, $4d$, or $5d$ transition-metal elements, with at most 14 atoms in the primitive cell. 
Nonmagnetic semiconductors were identified from the MC3D metadata by requiring both the total magnetization and the absolute magnetization to be zero. 
We further selected compounds with database PBEsol \cite{perdew2008} band gaps between 0.1 and 3.0 eV.
These criteria define the initial set of nonmagnetic semiconductors used in our workflow.

To screen for light-induced magnetism, we performed constrained DFT calculations for a photoexcited state with a fixed carrier density of 0.2 electrons per unit cell. 
Spin-polarized calculations were initialized from magnetic configurations generated within the AiiDA \cite{pizzi2016,huber2020} workflow. 
A material was identified as a candidate if the converged photoexcited state exhibited an absolute magnetic moment larger than 0.5 $\mu_B$ per photoexcited electron. 
The magnetic order of the photoexcited state was classified from the total and absolute magnetic moments per photoexcited electron: ferromagnetic (FM) states satisfy 
$|M_{abs}-M_{tot}|<0.1\mu_B/e$, antiferromagnetic (AFM) states satisfy $|M_{tot}|<0.1 \mu_B/e$ and the remaining magnetic states were classified as ferrimagnetic (FR).
Structural motifs were classified manually based on the local coordination environment of the transition-metal site.
For each chemical group, the reported average induced moments were computed over the full candidate set, while groups containing fewer than five materials were excluded.

    To improve the screening-level PBE gap estimates, corrected band gaps were assigned hierarchically: MBJ \cite{choudharyComputationalScreeningHighperformance2018a} values were used when available, otherwise all-electron HSE06 \cite{nair2025} values were used, and remaining compounds were assigned machine-learning-predicted gaps trained on MBJ data \cite{choudhary2021}. 
These corrected gaps were used for the statistical analysis of optical gap windows discussed in the main text.

\subsection{Computational details}
The screening calculations were carried out with Quantum ESPRESSO \cite{giannozzi2009,giannozzi2017} using the PBEsol \cite{perdew2008} exchange-correlation functional and the SSSP precision pseudopotential library \cite{prandini2018}. 
Plane-wave cutoffs were set automatically by the AiiDA \cite{pizzi2016,huber2020} workflow with protocol="stringent". 
The Brillouin zone was sampled using automatically generated k-point meshes with a target distance of 0.10 1/\AA, and the self-consistency threshold was fixed at $10^{-8}$. 

The photoexcited state was described within constrained DFT, which imposes equal electron and hole populations corresponding to a quasi-equilibrium electron-hole plasma. 
A fixed photoexcited carrier density of 0.2 e/u.c. was used throughout the screening. 
Electronic occupations were treated with Fermi-Dirac smearing of 0.0002 Ry, and spin-polarized calculations were used. 
The full workflow was implemented in AiiDA.

\section*{Acknowledgements}
We acknowledge fruitful discussions with Salvatore Lubreto and Giovanni Marini.  We acknowledge the CINECA award under the ISCRA initiative (project IscrC\_Uf--DynFP), for the availability of high performance computing resources and support. Funded by the European Union (ERC, DELIGHT, 101052708). Views and opinions expressed are however those of the author(s) only and do not necessarily reflect those of the European Union or the European Research Council. Neither the European Union nor the granting authority can be held responsible for them.

\bibliography{magnetization}
\end{document}